# Electric-Field Guided Precision Manipulation of Catalytic Nanomotors for Cargo Delivery and Powering Nanoelectromechanical Devices

Jianhe Guo[1], Jeremie June Gallegos[2], Ashley Robyn Tom[2], and Donglei Fan[1, 2*]

1 Materials Science and Engineering Program, the University of Texas at Austin, Austin, TX 78712, USA

2 Department of Mechanical Engineering, the University of Texas at Austin, Austin, TX 78712, USA

**ABSTRACT:** We report a controllable and precision approach in manipulating catalytic nanomotors by strategically applied electric (*E*-) fields in three dimensions (3-D). With the high controllability, the catalytic nanomotors have demonstrated new versatility in capturing, delivering, and releasing of cargos to designated locations as well as in-situ integration with nanomechanical devices (NEMS) to chemically power the actuation. With combined AC and DC *E*-fields, catalytic nanomotors can be accurately aligned by the AC *E*-fields and instantly change their speeds by the DC *E*-fields. Within the 3-D orthogonal microelectrode sets, the in-plane transport of catalytic nanomotors can be swiftly turned on and off, and these catalytic nanomotors can also move in the vertical direction. The interplaying nanoforces that govern the propulsion and alignment are investigated. The modeling of catalytic nanomotors proposed in previous works has been confirmed quantitatively here. Finally, the prowess of the precision manipulation of catalytic nanomotors by *E*-fields is demonstrated in two applications: the capture, transport, and release of cargos to pre-patterned microdocks, and the assembly of catalytic nanomotors on NEMS to power the continuous rotation. The innovative concepts and approaches reported in this work could further advance ideal applications of catalytic nanomotors, *e.g.* for assembling and powering nanomachines, nanorobots, and complex NEMS devices.

## INTRODUCTION

The integration of autonomous inorganic micro/nanomotors as components of micro/nanomachines with high precision and versatility to power their operations is a critical step in realizing the ideal nanofactories and nanorobots, which could revolutionize modern lives.[1-9] Catalytic micro/nanomotors that convert chemical energy into mechanical motions, are one of the most widely exploited autonomous nanomotors.[10-18] Billions of catalytic nanomotors can be facilely fabricated by using a variety of techniques, such as electrodeposition into nanoporous templates and electron beam deposition on monolayer nanospheres.[1-2] They self-propel by harvesting chemical energies from fuels in suspension, such as hydrogen peroxide.[18-19] Recently, substantial research efforts have been focused on strategically designing and fabricating catalytic nanomotors with an array of compositions and geometries, such as bimetallic nanorods,[18-20] catalytic microtubes,[21-23] and Janus particles[24-27]. The efforts lead towards dramatic improvement of propulsion speeds up to hundreds of μm/sec (or 100 body lengths per second),[19-21] and readiness in harnessing energy from a variety of fuels, such as hydrazine,[28] urea[29-30] and even pure water[13, 26]. More importantly, vast applications of catalytic nanomotors have been demonstrated, such as on-chip cargo transport,[24, 31] drug delivery,[32-33] microchip repair,[34] nanolithography,[35] biomolecular sensing and in-vivo disease treatment.[36]

However, it remains challenging to align catalytic nanomotors with high precision and modulate their moving speeds facilely and instantly. The ability to achieve this could open unprecedented opportunities. Innovatively, magnetic fields have been exploited in guiding catalytic nanomotors, however, this strategy requires the integration of magnetic elements in the nanomotors and precise alignment of magnetic moments.[24, 31] Also, to generate magnetic forces, bulky electromagnets are often employed, which could be the bottleneck when developing portable nanomotor based devices. Acoustic tweezers have been used in guiding catalytic nanomotors to aggregate and disperse,[37] while, the resolution in manipulation is restricted by the large wavelength of acoustic waves. Besides controlling the orientation of catalytic nanomotors, it is of paramount importance to facilely tune their speed. Several unique approaches have been exploited to control locomotion speed of catalytic nanomotors. With the strong dependence on catalytic reactivity, the speed of catalytic micro/nanomotors can be tuned by localized stimuli, including fuel concentration,[18] temperature,[38] and light illumination.[26, 39-41] Furthermore, by applying electrical potentials to create chemical gradient,[42] or generating ultrasonic waves,[37, 43] the speed of micro/nanomotors can also be modulated. However, it remains difficult to realize both the guiding and speed tuning of catalytic nanomotors with high accuracy, facileness, and in an all-on-chip manner.

In this work, we report a unique approach for manipulating catalytic nanomotors with high precision and facileness. The work is based on strategically combined AC and DC $E$-fields, the so-called electric tweezers, applied via a 3-D orthogonal microelectrode setup.[44] Here, the DC $E$-field tunes the transport speed via electrophoretic and electroosmosis effects. The AC $E$-field guides the alignment independently via electric torques on the induced dipoles of nanomotors. By applying the combined AC and DC $E$-fields in 3-D, catalytic nanomotors can instantly align, transport along defined directions, start and stop, and change speeds on demand. The involved various nanoforces governing the motions are investigated. Leveraging the high precision in the alignment, the linear dependence of speed on the inverse of size of nanomotors ($1/l$) down to submicrometers, is experimentally determined, confirming previous theoretical predictions.[14] Finally, the manipulation of catalytic nanomotors by $E$-fields is demonstrated for two applications: the dynamic loading, transport, and unloading of micro-targets to pre-patterned microdocks; and assembling and integration of a catalytic nanomotor on a rotary NEMS to power its continuous operation.

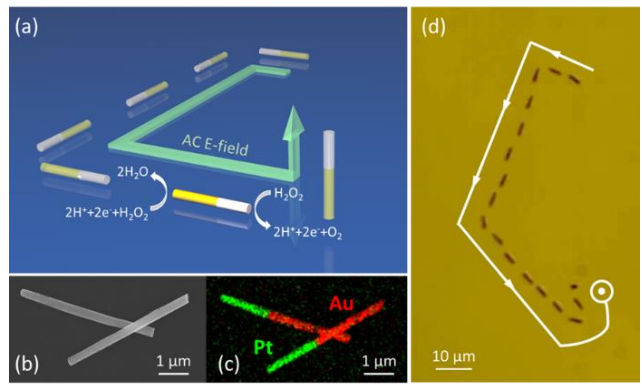

Figure 1. (a) Schematic diagram of 3D manipulations of Pt-Au catalytic nanomotors in $H_2O_2$ fuel with AC E-fields. (b) Scanning electron microscopy (SEM) and (c) Energy-dispersive X-ray spectroscopy (EDX) images of Pt-Au catalytic nanomotors (250 nm in diameter, 5 μm in length; consisting of 2-μm Pt segment and 3-μm Au segment). (d) Overlapped snapshots of a catalytic nanomotor guided by AC E-fields.

## RESULTS AND DISCUSSION

### *Guide, Start and Stop, and Modulate Speed of Catalytic Nanomotors*

The demonstrations of transport guidance and speed modulation of catalytic nanomotors by $E$-fields are carried out by using the classical Pt-based bimetallic nanorod motors as a model system [Fig. 1(a)]. Arrays of multi-segmented Pt-Au nanorod motors are synthesized with controlled lengths and diameters by electrodeposition into nanoporous templates in a three-electrode setup.[18, 45] The fabrication details are provided in the Supporting Information.

Scanning electron microscopy (SEM) and energy-dispersive X-ray spectroscopy (EDS) in Fig. 1(b)-(c) confirm the uniform cylindrical morphology, controlled size and composition of the Pt-Au catalytic nanomotors.

A software interfaced 3-D orthogonal microelectrode setup is designed and constructed for guiding the catalytic nanomotors as shown in Scheme Fig. 2. The set-up includes in-plane quadruple microelectrodes patterned on a glass slide and a pair of indium tin oxide (ITO) parallel electrodes assembled in the vertical direction. The bottom ITO electrode is fabricated on the opposite side of the quadruple microelectrodes on the glass (1 mm in thickness). A suspension reservoir is formed on top of the quadruple microelectrodes by a piece of polydimethylsiloxane (PDMS) elastomer (~1 mm in thickness) with a well of ~ 4 mm in diameter. The top of the well is sealed with a second ITO electrode for providing $E$-field in the vertical direction.

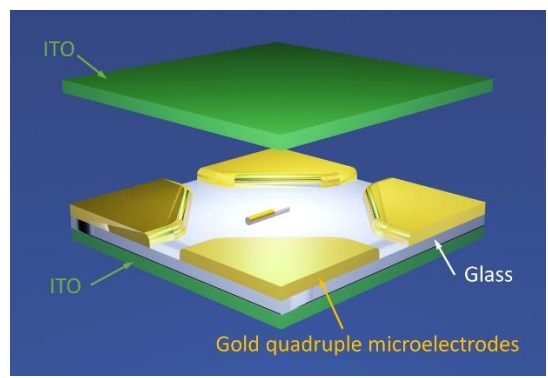

Figure 2. Scheme of 3-D orthogonal microelectrode setup.

Applying an AC $E$-field on the in-plane quadruple microelectrodes, the catalytic nanomotors can be instantly aligned along the direction of the AC $E$-field and move autonomously in the direction with the Pt segment as the front, as shown in the schematic diagram in Fig. 1(a), overlapped images in Fig. 1(d) and movie S1 (Supporting Information). We found that an AC peak-to-peak voltage of 10 V is sufficient to align the catalytic nanomotors and to guide their motions. When superimposing a DC $E$-field on the AC $E$-field, the speed of the nanomotors can increase or decrease instantaneously, and even reverse the moving direction. We characterize the manipulation in detail in the following to unveil the fundamental interactions between the nanomotors and $E$-fields.

First, we investigate the dependence of speed of the catalytic nanomotors on the concentration of hydrogen peroxide fuel ($H_2O_2$), both with and without $E$-fields. The average speed of the catalytic nanomotors is determined statistically from the behaviors of 10 nanomotors for 10 seconds (details in the Supporting Information). As shown in Fig. 3(a), regardless whether the $E$-field is applied or not, the speed of the nanomotors increases with the concentration of $H_2O_2$ and reaches a plateau, which can be attributed to the saturation of catalytic active sites on the nanomotors in high concentration $H_2O_2$ fuels.[18] Consistently, we find that the speed of these catalytic nanomotors under AC $E$- fields is always higher than those without AC *E*-fields. Furthermore, the speed of all nanomotors, regardless of their moving directions, increases when applying a uniform AC *E*-field. Neither the electroosmosis flows nor the induced dielectrophoretic forces could result in the observed behavior, so the dominating factor is the reduction of rotational Brownian motions of nanomotors due to the alignment by AC *E*-fields. Our experimental results and analysis support this understanding. First, we determine the speed of nanomotors as a function of AC frequency and voltage. It is found that the average speed of nanomotors monotonically increases with AC frequency before reaching a constant at around 500 KHz and 20V. At a fixed AC frequency, i.e. 500 KHz, the average speed also increases with AC voltage amplitude until reaching a constant at 30 V as shown in Fig. 3(b)-(c). The voltage dependence can be readily understood from the increase of electric torque ($\tau_e$) with applied *E*-field (*E*), which counters the rotational Brownian motions and thus enhances the degree of alignment of a nanomotor, given by:[46]

$$\tau_e = \frac{8\pi l r^2}{3} \cdot P \cdot E^2 \cdot sin2\theta \quad (1)$$

where *l* and *r* are the length and radius of nanomotor, respectively; $\theta$ is the angle between the long axis of nanomotor and the *E*-field; *P* is a value determined by the permittivity and conductivity of the medium and nanomotor as well as the AC frequency. We observe a leveling off of the moving speed when the voltage is adequately high, i.e. above 30 V, as shown in Fig. 3(b). It could be understood that when the electric torque is sufficiently high, the rotational Brownian motion is suppressed to an extent that the increase of speed with voltages is too small to be determined compared to the statistic distributions of speed of tested nanomotors.

By analyzing electric torques, we can also attribute the observed dependence of nanomotor speed on AC frequency [Fig. 3(c)] to the suppressed rotational Brownian motions. We experimentally determine electric torques as a function of angular positions when aligning nanomotors at 5 KHz to 50 MHz with Eq. (1). The $\tau_e$ exerted on the catalytic nanomotors at an angle of $\theta$ can be readily obtained from the angular velocity ($\omega$) versus angle ($\theta$) [Fig. S1, supporting information], since the viscous torque ($\tau_\eta$) instantly counters the electric torque ($\tau_e$) as given by $\tau_e = \tau_\eta$ in low Reynolds number environment. Here, the viscous drag torque $\tau_\eta$ on a rotating nanorod is calculated as follows:[47]

$$\tau_\eta = \frac{1}{3}\omega\pi\eta l^3 \frac{N^3-N}{N^3(\ln\frac{l}{Nr}+0.5)} = K_1 \omega \ (N\ m) \quad (2)$$

where $\eta$ is the viscous coefficient of suspension medium and *N* is the number of nanorod segments, taken as 2 in this calculation. For the nanomotors used in our experiments, the constant $K_1$ is determined as $2.50\times10^{-20}$. Now, with $\tau_e$ determined at different $\theta$ from Eq. (1) and (2), we can readily obtain the coefficient of $\tau_e$ at different AC frequencies as shown in Fig. 3(c). It can be readily found that the dependence of nanomotor speed on AC frequency well matches the dependence of electric torque applied on nanomotors versus AC frequency. It well supports the key contribution of alignment by AC *E*-fields to the observed enhanced speed of nanomotors.

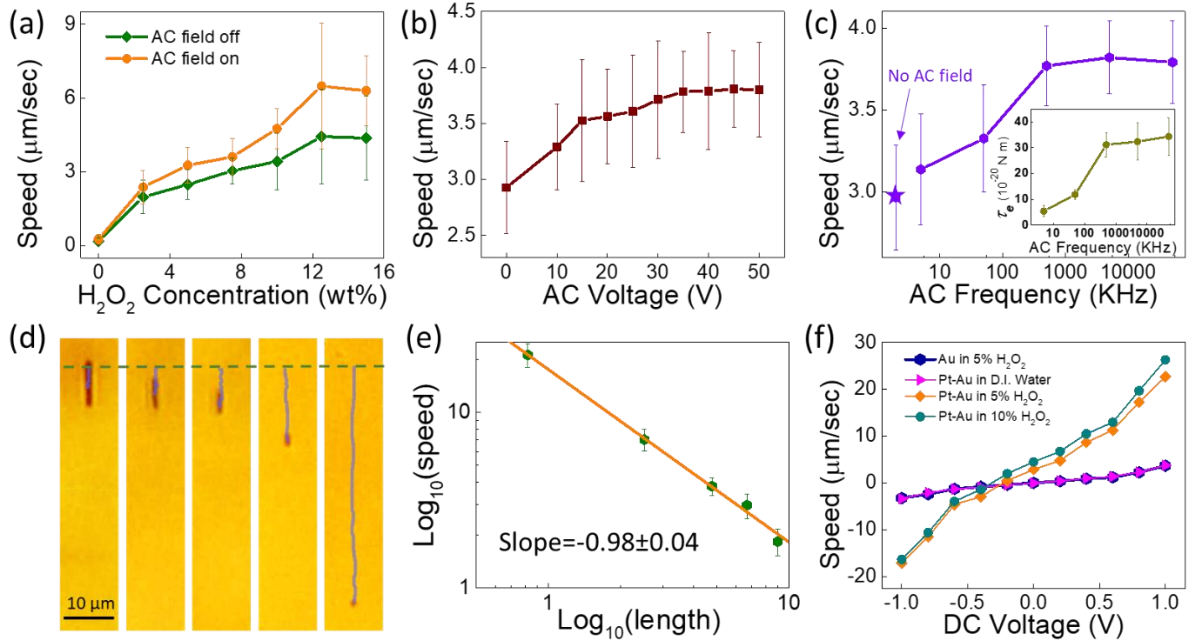

**Figure 3.** (a) Speed of catalytic nanomotors versus concentration of $H_2O_2$ with/without AC *E*-fields (5 MHz, 15 V). (b) Speed versus peak-to-peak voltage of AC *E*-fields (500KHz) in 7.5 wt% $H_2O_2$ solution. (c) Speed versus frequency of AC *E*-fields (20 V) in 7.5 wt% $H_2O_2$ solution. Inset: electric torque versus frequency of AC *E*-fields. (d) Trace of catalytic nanomotors of different length in 2 seconds. The lengths of nanomotors (from top to bottom) are 0.82, 2.50, 4.78, 6.68 and 8.96 μm, respectively. The green dashed line shows the starting position. (e) Log-log plot of speed versus length of nanomotors. The slope is -0.98. (f) Speed of Au nanorods and Pt-Au catalytic nanomotors of same dimensions versus DC *E*-fields in different suspension mediums.

With the uniform AC *E*-fields, we not only improve the alignment and speed of nanomotors as discussed as above, but also successfully guided catalytic nanomotors in the vertical direction in a facile and all-on-chip manner. Nanomotors can move along prescribed trajectories in the 2-D X-Y plane, start and stop on demand, and even move vertically as shown in Fig. 1(a), Movie S1 and S2 in Supporting Information. For nanomotors moving in the X-Y 2-D plane, it is known that the catalytic driving force balances with the drag force, given by:[48]

$$F_{catlytic\ drive} = F_{drag} = \frac{2\pi\eta l}{\ln\frac{l}{r}-0.5}v = K_2\ v\ (N) \qquad (3)$$

where $v$ is the velocity of catalytic nanomotors and the geometric factor $K_2 = 8.77\times10^{-9}$ for the tested nanomotors (5 μm in length and 250 nm in diameter). Note that the value of viscous coefficient $\eta$ of suspension medium is estimated from that of pure water at room temperature, knowing that the change of viscous coefficient is less than 3% when the concentration of $H_2O_2$ is less than 10 wt%.[49] In the vertical direction, a catalytic nanorod made of Pt(2-μm)-Au(3-μm) experiences gravitational and buoyant forces of $4.85\times10^{-14} N$ and $0.24\times10^{-14} N$, respectively. Therefore, from Eq. (3) and the experimental results of velocity versus fuel concentration ($C_{H_2O_2}$) shown in Fig. 3(a), we can readily determine that the catalytic driving force of nanomotors is sufficiently high to realize the propulsion of the nanomotor in the vertical direction when the concentration of fuel ($C_{H_2O_2}$) is above 10 wt%. This analysis is validated by experimental study as shown in movie S2 in the Supporting Information. When $C_{H_2O_2}$ is 12.5 wt%, where a driving force of $5.69\times10^{-14} N$ is determined by calculation, the catalytic nanomotors indeed aligned and transported vertically, overcoming gravitation forces, and gradually disappeared from the view. While at a lower fuel concentration, *e.g.* when $C_{H_2O_2}$ is 7.5 wt%, a vertically applied AC *E*-field can instantly align nanomotors vertically and stop the motions in 2-D planes. Here the driving forces due to catalytic reactions, calculated as $3.17\times10^{-14} N$, is not sufficient to propel the motors vertically. Both the demonstrations of 3-D manipulation at high fuel concentrations and "on/off" control of 2-D motion at low fuel concentrations offer considerable promise for versatile operations of these catalytic nanomotors, opening many opportunities for applications. Such a strategy could also be applied for controlling photocatalytic nanomotors[26, 40] and self-propelled enzyme nanomotors[12, 30].

The high precision in the alignment of nanomotors by AC *E*-fields offers new opportunities in studying and understanding the working mechanism of the self-propelled catalytic nanomotors. In a recent model of a catalytic nanomotor made of bimetallic nanorods, the velocity ($v$) due to self-electrophoresis is given by:[14]

$$v \sim \frac{\zeta_p\varepsilon\varepsilon_0}{\eta}E_{int} \sim \frac{\zeta_p\varepsilon\varepsilon_0\Delta\phi}{\eta l} \qquad (4)$$

where $\varepsilon$ is the relative dielectric constant of the suspension medium and $\varepsilon_0$ is the dielectric permittivity of free space. The self-generated *E*-field $E_{int}$ is approximated as the potential drop $\Delta\phi$ divided by the length of the nanorods $l$ and $\Delta\phi$ is determined by the chemical potential of the two metal segments of a nanomotor, which is considered as a constant with different lengths; $\zeta_p$ is Zeta potential of a nanomotor, independent of the size of the nanomotor (Fig. S3). Therefore, we can readily find that the self-electrophoretic velocity ($v$) of a nanomotor should be inversely proportional to its length ($l$) according to Eq. (4) given by the model. Although in previous studies, it has been shown that the smaller the catalytic nanomotors, the higher the moving speed, quantitative study has yet been done on the catalytic nanomotors as predicted by the model.[14, 50] To experimentally determine the dependence of speed on size of nanomotors, the key is to precisely align nanomotors with controlled moving trajectories and to suppress noises due to Brownian motion. As shown in the above studies, AC *E*-field alignment provides a facile and effective tool, where the alignment is determined by the overall shape anisotropy of a nanomotor. Indeed, with AC *E*-fields, we can facilely synchronously align catalytic nanomotors with sizes ranging from 0.82 to 8.96 μm (details in Fig. S4). When suspending in $H_2O_2$ solutions, the shorter nanomotors exhibit higher moving speeds compared to longer ones [Fig. 3(d)]. The speed and length of nanomotors follows an inverse proportional relationship as shown in the log-log plot with a slope of -0.98±0.04 in Fig. 3(e). It indicates that the power law dependence of speed on size of nanomotors is approximately -1, which provides quantitative proof of the working mechanism proposed previously for nanomotors in the size range of 0.82 to 8.96 μm. This result points towards the great advantages of catalytic nanomotors in achieving ultrahigh propulsion speed when made into ultra-small dimensions.[23, 51-52]

Next, we exploit the effect of DC *E*-fields on nanomotors. The speed of the Pt-Au catalytic nanomotors due to DC *E*-fields are tested in pure water and 5-10 wt% $H_2O_2$ fuels. Gold (Au) nanorods of the same dimension are fabricated and tested at the same conditions for control experiments. AC *E*-fields are superimposed to align the nanomotors in their long axis direction when applying DC voltages of -1 to 1 V on 500 μm-gapped microelectrodes. The speed of catalytic nanomotors linearly depends on DC *E*-fields and can be controlled to instantly increase, decrease and even reverse directions, depending on the magnitude and direction of DC *E*-fields as shown in Fig. 3(f) and movie S3. A DC voltage as low as 1 V can lead to a velocity change of ~20 μm/sec of the catalytic nanomotors, which is effective in modulating speed of nanomotors for various applications [Fig. 3(f)].

We also observe that the speeds of both catalytic nanomotors in fuel solutions and those in control experiments always increase in the direction of DC *E*-fields. While the Zeta potentials of both Pt-Au nanomotors and Au nanorods are negative as shown in the measurements in Fig. S2. Furthermore, although the speed of the catalytic nanomotors show much stronger responses to the DC *E*-field, the value of their Zeta potential in fuel solutions is significantly lower than those of control samples, *i.e.* for Pt-Au nanomotors in 5% $H_2O_2$ fuel, the Zeta potential of -8.76 mV is around 1/4 of those of control samples.

The above phenomena cannot be explained simply by electrophoresis (EP). Nanoentities, such as nanomotors, experience an electrophoretic force in the presence of an external DC *E*-field due to the formation of an electric double layer at the solid/liquid interface with opposite local charges. The speed of nanoentities ($v_{EP}$) due to electrophoretic force is proportional to the dielectric constant of suspension medium ($\varepsilon$), Zeta potential of nanomotor ($\zeta$), and DC *E*-field ($E_{ext}$), given by:[53]

$$\boldsymbol{v}_{EP} = \frac{\varepsilon \varepsilon_0 \zeta}{\eta} \boldsymbol{E}_{ext} \quad (5)$$

Given that the Zeta potentials of the catalytic nanomotors and control samples are all negative, if only because of the electrophoretic effect, the moving speed should increase in the opposite direction of the DC *E*-field and scale with the magnitude of the Zeta potential. However, the experiments show the behaviors of nanomotors opposite to this analysis. In-depth study shows that in addition to electrophoretic (EO) forces, electroosmosis flows generated on the surface of glass substrates are important as well, which is given by:[53]

$$\boldsymbol{v}_{EO} = -\frac{\varepsilon \varepsilon_0 \zeta_s}{\eta} \boldsymbol{E}_{ext} \quad (6)$$

The Zeta potential of glass substrates ($\zeta_s$) is negative and can reach -80 mV.[54] As a result, the direction of electroosmosis flow on glass substrates is in the same direction of DC *E*-field according to Eq. (6), opposite to that of the electrophoretic effect on nanomotors in Eq. (5). Therefore, the speed of a nanomotor ($v_E$) in DC *E*-field is governed by the combined effects of electrophoretic force on the nanomotor ($v_{EP}$) in Eq. (5) and electroosmosis flows on the glass substrate ($v_{EO}$) in Eq. (6). Considering the zeta potentials and the moving direction of nanomotors, it can be found that the electroosmosis flows ($v_{EO}$) dominate the motions of nanomotors. Therefore, lower absolute values of the negative zeta potential of the nanomotors lead to higher speed modulations by the DC *E*-field. This well agrees with the experimental results in Fig. 3(f), where the speed of catalytic nanomotors can be strongly tuned by the external DC *E*-field (Supporting Information).

With the demonstration and understanding of the prowess of AC and DC *E*-fields for manipulation of catalytic nanomotors with high facileness, precision, and efficiency, we exploited two applications of these catalytic nanomotors: targeted cargo delivery and assembling of catalytic nanomotors for powering rotary NEMS devices.

*Targeted Cargo Delivery*

In this demonstration [Fig. 4(a) and movie S4], Au nanorods (250 nm in diameter and 3.6 μm in length), fabricated by electrodeposition into nanoporous templates, serve as cargos. They are mixed with Pt-Au catalytic nanomotors in 7.5 wt% $H_2O_2$ fuel solution. Without AC *E*-field, the catalytic nanomotors move randomly, while the Au cargos only exhibit weak Brownian motions. By applying a uniform AC *E*-field, both nanomotors and cargo nanorods are aligned. The nanomotors move in the alignment direction. While, the cargo stays essentially at the original location during the transport of the nanomotor guided by the AC *E*-field as shown in Fig. 4(b)-(c). When the nanomotor are close to the cargo, the induced *E*-fields can readily assemble them tip to tip. Here, the interaction between the induced dipoles of nanomotor and cargo ensures a simple procedure to upload the cargo on the nanomotor and the assembly is robust during the transport in AC *E*-fields. Next, guided by the AC *E*-field, the nanomotor propels the cargo to a patterned metallic microdock as shown in Fig. 4(c)-(d). When the cargo is in the vicinity of the microdock, it can rapidly anchor to the edge of the metallic microdock due to the interaction between induced electric dipoles. Then the nanomotor can start its next journey to seek other targets as shown in Fig. 4(e)-(f). Here, the on-demand release of cargos from the catalytic nanomotors are facilely accomplished by turning off the AC *E*-field swiftly, *e.g.* for a few seconds [Fig. 4(e)]. The above process is further determined quantitatively by analyzing the instantaneous speed of nanomotors as shown in Fig. 4(g). When the nanomotor moves towards the cargo, the average speed is around 3.9 μm/sec. The speed swiftly increases to 5.7 μm/sec when it gets close and attaches to the cargo due to the strong mutual attraction. After capturing the cargo, the speed decreases to 2.2 μm/sec due to the load, which agrees with our estimation of 2.6 μm/sec from Eq. (3). After releasing the cargo, the nanomotor immediately restores to a speed up to 3.3 μm/sec, which is slightly lower than the speed of 3.6 μm/sec before loading the cargo. Note that at this moment the nanomotor is not aligned by the AC *E*-field which could account for the lowered speed.

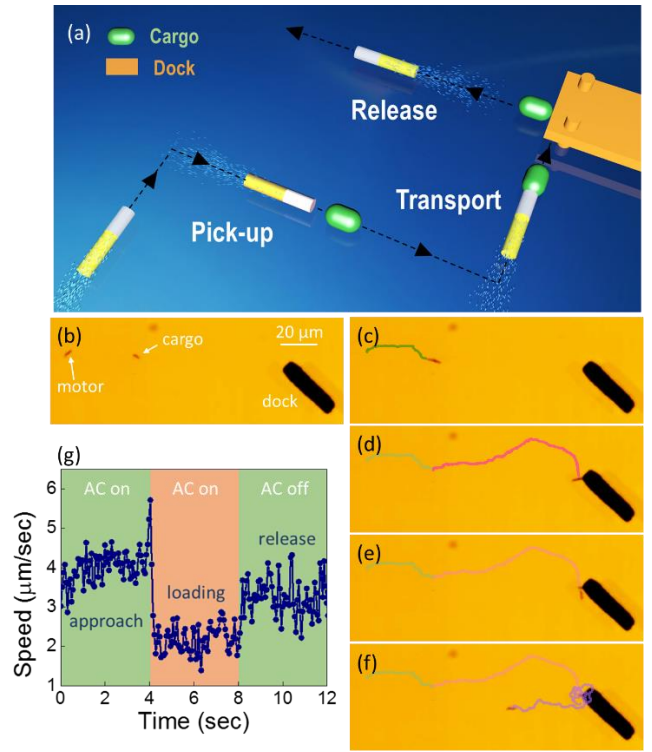

Figure 4. (a) Scheme of targeted cargo delivery. (b-f) Optical microscopy images show the dynamic process of a Pt-Au catalytic nanomotor when picking up, pushing, and delivering a nanorod cargo to the designated microdock. (g) Speed of the nanomotor in the cargo delivery process.

The above targeted cargo delivery is the first demonstration of *E*-field enabled loading and unloading of cargo by nanomotors, showing simplicity and reliability for diverse lab-on-a-chip applications.

### Powering Rotary NEMS by Precision Assembling Catalytic Nanomotors

Next, leveraging the precision guidance of catalytic nanomotors with *E*-fields, we designed and assembled a unique type of chemically powered rotary NEMS by integrating catalytic nanomotors with electric manipulation. Firstly, a rotary NEMS is assembled by *E*-fields following previous reports as illustrated in Fig. 5(a).[47, 55] It consists of a multi-segmented Au/Ni nanorod (250 nm in diameter and 8 μm in length) and a patterned nanomagnet (500 nm in diameter) serving as the rotor and bearing, respectively. Different from previous research, the rotor is designed to consist of three Ni segments embedded in the Au nanorod. The structure is Au(200 nm)/Ni(50 nm)/Au(3.5 μm)/Ni(500 nm)/Au(3.5 nm)/Ni(50 nm)/Au(200 nm) as shown in Fig. 5(b). The Ni segment in the center of the rotor is used to attach the rotary NEMS on the patterned magnetic bearing. The two Ni segments next to the tips aim to anchor the catalytic nanomotors. The patterned nanomagnet has a tri-layer stack of Ti (60 nm)/Ni (80 nm)/Cr (6 nm). By using the electric tweezers, the nanorotor is transported and attached atop of the pattern magnetic bearing, where the magnetic interaction between the Ni segment in the rotor and Ni layer in the magnetic bearing fixes the position of the rotary NEMS while still allows its rotation.[47, 55] Next, we fabricated catalytic nanomotors made of Pt-AgAu nanowires. Here, we replaced the previously used Au segments with Ag(~80 wt%)/Au(~20 wt%) alloys in the Pt-based catalytic nanomotors to substantially increase the output power and speed due to the enhanced potential differences between the two segments as shown in Fig. 5(c-d).[19] The catalytic nanomotors move at a speed up to ~30 μm/sec in 5 wt% $H_2O_2$ solution. From the speed, we can estimate the driving force as high as 0.26 piconewtons (pN). With the guidance of the AC *E*-field, the catalytic nanomotor can be efficiently maneuvered towards the rotary NEMS and then be assembled at one end of the rotor by magnetic attraction as shown in Fig. 5(a). After assembling, the nanomotor instantly drives the NEMS into continuous rotation, all powered by the chemical energy harvested from the $H_2O_2$ fuels. The driving torque can be readily determined as $1.04 \times 10^{-18} N \cdot m$, which is sufficient to overcome the friction and magnetic torques between the rotor and bearing as shown in Fig. 5(e-g) and movie S5. Here, we can observe the effects of the angle-dependent magnetic force and torque that result in the oscillation of rotation speeds of the rotor with a 360° periodicity as a function of the angular position, agreeing with previous works.[47, 56] The average rotation speed of the NEMS device is determined as 0.64 rad/sec.

This work demonstrates a new approach in integrating powering components to drive nanomechanical devices, which could be applied to assemble various functional components of nanorobots and complicated NEMS/MEMS devices with far-reaching impact in electronics and biomedical research.

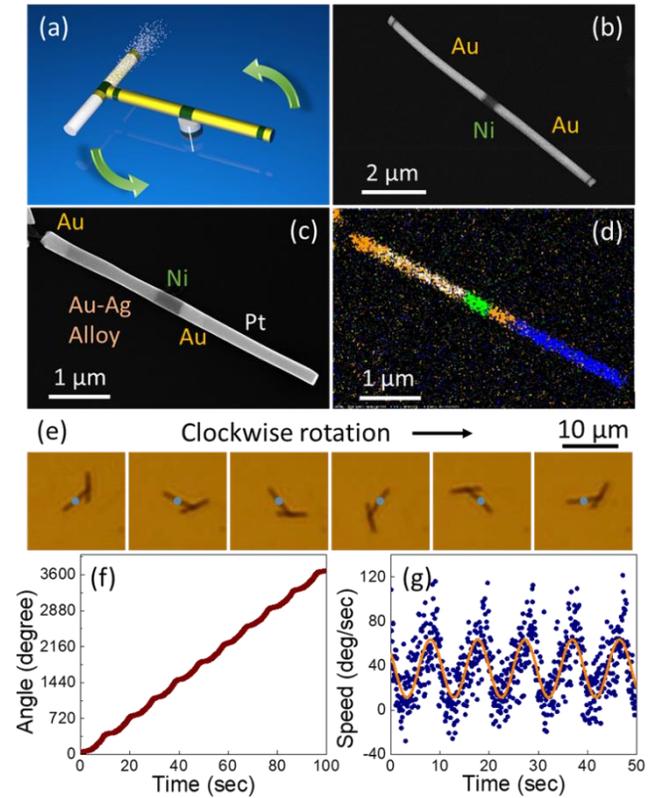

Figure 5. Assembling of catalytic a nanomotor for powering NEMS device into continuous rotation. (a) Schematic diagram of the rotary NEMS device with catalytic nanomotor assembled as the powering component. (b) SEM images of the multi-segment nanowire rotor. The segments from left to right are 200 nm Au, 50 nm Ni, 3.5 μm Au, 500 nm Ni, 3.5 μm Au, 50 nm Ni and 200 nm Au. (c-d) SEM and EDX images of the Pt-Ag/Au catalytic nanomotor. The segments from left to right are 400 nm Au, 1.7 μm Ag-Au alloy, 500 nm Ni, 400 nm Au, and 2 μm Pt. (e) Snapshots of the rotary NEMS device taken every 2 seconds. (f) Rotation angle versus time. (g) Rotation speed versus time.

## CONCLUSIONS

In summary, we report a new, controllable and precision approach to guide and modulation propulsions of catalytic nanomotors with *E*-fields, which allows for a motion control in 2-D and 3-D. The fundamental interactions involved in the electric manipulation have been investigated. Leveraging the precision of manipulation, we experimentally confirmed the inverse linear dependence of speed and size of catalytic nanomotors, supporting previous modeling. For applications, the manipulation strategy provides facile operations of these catalytic nanomotors to realize cargo capture, transport, and delivery to designated microdock. The prowess of the manipulation is also demonstrated by assembling catalytic motors as powering component of rotary NEMS. This innovative approach could be applied to construct various nanorobots and functional NEMS/MEMS devices for diverse tasks in electronics and biomedical research.

## ASSOCIATED CONTENT

**Supporting Information**.
This material is available free of charge via the Internet at http://pubs.acs.org.
Experimental details, velocity analysis, torque calculation and additional Zeta potential and SEM characterization of catalytic nanomotors. (PDF)
Movie S1. Random movements of conventional catalytic nanomotor versus directional motions of E-field guided catalytic nanomotor (2× frame rate).
Movie S2. Motions of catalytic nanomotors in the vertical direction with the alignment of the AC E-fields. The fuels have concentrations of 7.5 wt% and 12.5% H2O2, respectively.
Movie S3. Speed modulation of a catalytic nanomotor (2× frame rate).
Movie S4. Targeted cargo delivery by a catalytic nanomotor (5× frame rate).
Movie S5. Rotary NEMS device powered a catalytic nanomotor (5× frame rate).

## AUTHOR INFORMATION

Corresponding Author
* dfan@austin.utexas.edu
Notes
The authors declare no competing financial interest.

## ACKNOWLEDGMENT

This research is supported by the Welch Foundation (F-1734) and National Science Foundation (CMMI 1150767 and EECS 1710922). We appreciate the idea from Prof. Tom Mallouk in studying the relationship of speed and size of the catalytic nanomotors. We also thank the discussion with Prof. Peer Fischer on this work.

Table of Contents (TOC)

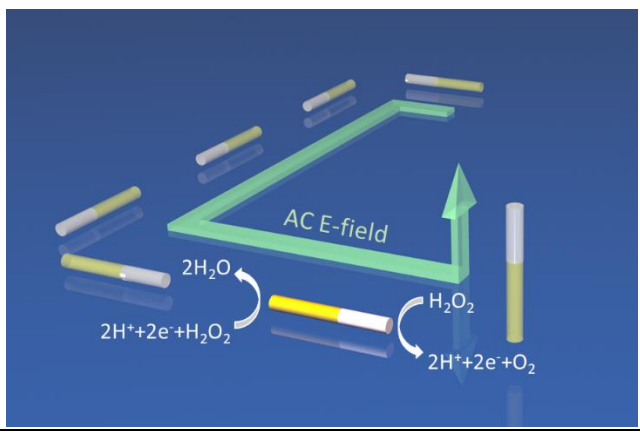